\documentclass[conference]{IEEEtran}
\IEEEoverridecommandlockouts

\usepackage{cite}
\usepackage{amsmath,amssymb,amsfonts}
\usepackage{algorithmic}
\usepackage{graphicx}
\usepackage{textcomp}
\usepackage{xcolor}
\def\BibTeX{{\rm B\kern-.05em{\sc i\kern-.025em b}\kern-.08em
    T\kern-.1667em\lower.7ex\hbox{E}\kern-.125emX}}
\usepackage{multirow}

\newcommand{\p}{\ensuremath{\mathbf{p}}}

\begin{document}

\title{
A Neural Model for Contextual Biasing Score Learning and Filtering
}

\author{\IEEEauthorblockN{Wanting Huang \hspace{5em} Weiran Wang}
\IEEEauthorblockA{\textit{Computer Science Department} \\
\textit{University of Iowa}\\
Iowa City, USA \\
wanting-huang@uiowa.edu \hspace{5em} weiran-wang@uiowa.edu}
}

\maketitle

\begin{abstract}
Contextual biasing improves automatic speech recognition (ASR) by integrating external knowledge, such as user-specific phrases or entities, during decoding. In this work, we use an attention-based biasing decoder to produce scores for candidate phrases based on acoustic information extracted by an ASR encoder, which can be used to filter out unlikely phrases and to calculate bonus for shallow-fusion biasing. We introduce a per-token discriminative objective that encourages higher scores for ground-truth phrases while suppressing distractors.  Experiments on the Librispeech biasing benchmark show that our method effectively filters out majority of the candidate phrases, and significantly improves recognition accuracy under different biasing conditions when the scores are used in shallow fusion biasing. Our approach is modular and can be used with any ASR system, and the filtering mechanism can potentially boost performance of other biasing methods. 
\end{abstract}


\section{Introduction}
\label{sec:intro}

In recent years, Automatic Speech Recognition (ASR) has made significant strides, largely thanks to the rise of end-to-end neural (E2E) models such as Connectionist Temporal Classification (CTC,~\cite{graves2006connectionist}), Recurrent Neural Network Transducer (RNN-T,~\cite{Graves_12a}), and Attention-based Encoder–Decoder (AED,~\cite{Chorowski15a,chan2016listen}). These models have achieved impressive results on general benchmarks. However, they still face notable challenges when it comes to recognizing rare or domain-specific terms—such as proper nouns, technical jargon, or personalized content—that appear infrequently in the training data. This shortcoming becomes especially evident in real-world scenarios like virtual assistants or personalized transcription services, where users naturally expect the system to understand context-specific vocabulary.

To bridge this gap, contextual biasing has emerged as an effective strategy. It works by injecting a curated list of task- or user-specific phrases into the ASR decoding process, helping the model give preference to relevant terms. 
Biasing methods are broadly divided into two categories. The first includes inference-based biasing methods, which augment the scores of hypotheses containing specified biasing phrases during beam search. These methods typically operate without learnable parameters and do not necessitate a training regimen. The second category encompasses model-based biasing methods, characterized by their direct incorporation of contextual information into the ASR architecture during the training phase.
We provide a detailed review of both approaches in Section~\ref{sec:related}.
Both inference-based and model-based approaches have their drawbacks. Inference-based methods offer modularity and flexibility, easily integrating into any ASR system, but they struggle to differentiate the probability of various phrases. Conversely, model-based methods can lead to increased architectural complexity. A shared limitation is that both incur substantial computational costs when dealing with a vast number of potential phrases.

In this paper, we present a new approach: we train a biasing decoder on candidate biasing phrases to produce a score for each phrase which reflects the likelihood of the phrase appearing in audio, based on acoustic information extracted by an ASR encoder.
Different from most model-based approaches that perform cross-attention to phrase embeddings, the biasing decoder works in the same way as an autoregressive attention decoder (and thus our method can be interpreted as an audio-based language model), except that it only produces the likelihood for candidate phrases instead of the full ASR transcript. 
Borrowing ideas from standard ASR, our training objective for the biasing decoder ensures it learns a discriminative per-token score. The learned score is used to filter out majority of unlikely phrases, and can also be used to calculate bonus scores in inference-based biasing.
This design allows for plug-and-play integration with existing systems, without requiring any changes to the underlying ASR components.



We validate our method on the public Librispeech biasing benchmark~\cite{le2021contextual}. Paired with shallow-fusion biasing~\cite{wang2024contextual}, our method consistently achieves significant reductions in Word Error Rate (WER) with different number of phrases, and compares favorably with prior works. In the most challenging case of 2000 distractors, our method keeps less than 1\% of the phrases during search, reduces the test-clean ASR WER from 2.7\% to 2.1\%, and test other from 6.3\% to 5.0\%, and achieves over 50\% relative WER reduction on infrequent words.    

\section{Related Work}
\label{sec:related}

Contextual biasing has emerged as a key research focus in automatic speech recognition (ASR), particularly for improving the recognition of rare or domain-specific terms. Broadly, contextual biasing approaches can be grouped into inference-based and model-based methods.

\subsection{Inference-Based Biasing }
\label{sec:shallow-fusion}

Inference-based methods incorporate biasing contextual information primarily during the decoding process, often without modifying the ASR model itself. 
This approach offers modularity and flexibility. 
Predating the widespread adoption of E2E models, researchers focused on injecting biasing contexts to boost decoding scores for specific words or phrases\cite{apetar,Hall2015CompositionbasedOR}. Typically, this involved constructing a compact search graph for the phrases and composing it with the normal ASR search graph, in the Weighted Finite State Transducer (WFST) framework~\cite{mohri2002weighted}.
Weights along the biasing graph edges would then add bonus scores to hypotheses matching these phrases. 
For on-device speech recognition with an E2E model, namely RNN-T, \cite{zhao2019shallowfusion} adapted this approach by incorporating bonuses at the subword level.
When the bonus scores are incorporated before beam pruning, this method is known as \emph{shallow fusion} because it's similar to how external language model scores are used during inference. 

\noindent\textbf{Implementation}
In this work, we will be testing our method with the inference-based approach. We use the GPU-friendly implementation of shallow-fusion biasing proposed by~\cite{wang2024contextual}, which potentially facilitates better parallelization. 
Before search starts, a partial match table is built to provide the index to backup to when encountering a token mismatch for each phrase. During search, we maintain for each hypothesis the partial matching lengths with each biasing phrase. These are essentially the ``search state'' of shallow fusion, and the transition between states follows the pre-built partial match table.
In this approach, a per-token bonus is added to a token expansion during beam search, if this token extends the matching into a phrase in the biasing list, and when the token expansion stops matching into any phrase the accumulated bonus is canceled.
In~\cite{wang2024contextual}, the per-token bonus is a user parameter tuned on the development set. 
The goal of our work is to learn a neural model that provides a discriminative score.

\subsection{Model-Based Biasing}

Model-based methods embed a biasing component directly into the E2E ASR model's architecture or training process. To accept biasing contexts as an additional input, the cross-attention mechanism~\cite{Vaswani_17a} is often employed to condition the model's outputs on these contexts. With many architecture variants, the summarized contextual information is then propagated to the rest of the ASR model to influence predictions ~\cite{pundak2018deep,Jain2020ContextualRF,munkhdalai2021fast,chang2021automatic,han2022improving,sathyendra2022contextual,meng2024text}.



More recently, using contextual adapters~\cite{sathyendra2022contextual} for both encoder and predictor in the RNN-T model, \cite{tang2024improving} proposed a guided attention method to enforce the cross-attention weights to reflect the existence of a phrase at each audio frame and every output token, with additional cross-entropy or CTC loss.
\cite{Sudo2024} used a bias decoder to predict, for each output token, the index of phrase it belongs to from a phrase list (and a ``no-bias'' phrase is used when the token does not belong to any given phrase). They extracted an embedding for each phrase in order to make phrase level prediction, and the bias decoder was trained with a cross-entropy loss. At inference time, they would boost the score of a token if it belongs to a phrase predicted by the bias decoder. In other words, they performed ranking and filtering of phrases on the fly during decoding.
\cite{Wang2024ContextBalanced} identified issues with context adapters and proposed a balanced learning objective to guide attention mechanisms more effectively, particularly for rare phrases.

For deeper integration of contextual information into the joint RNN-T/CTC model,~\cite{Shakeel2024} proposed to use biasing losses on intermediate representations from the audio encoder. The intermediate biasing losses are computed with an CTC decoder, and the target sequences are obtained from the ASR transcripts by keeping only the biasing phrase and replacing the rest with ``no-bias'' tokens.  
In~\cite{Huang2024}, the authors explored methods for enhancing contextualization within an RNN-T model. Their approach involved introducing biasing lists at the intermediate encoder layers and employing text perturbations, specifically alternative spellings, to compel the model to utilize contextual information
\cite{Fang2025} used a multi-label synchronous output CTC loss to enhance the synchronization between the ASR output by CTC and the bias outputs by another CTC, and improved phrase-level contextual representation. 

Different from most model-based approaches, \cite{le2021deep} and~\cite{sun2021tree} incorporated tree-based symbolic representation of biasing list into E2E model's forward function. Intuitively, this approach is one step closer to shallow fusion.

With the advent of large language model (LLM)-based ASR systems~\cite{lakomkin2023endtoends,wang2023slm,chen2024salm,kong2024audio,hori2025delayedfusion}, multiple works have tried to enhance their biasing capability.
Besides cross-attention based biasing, there exists another two major approches.
One approach is text-based rescoring or hypothesis editing. 
~\cite{Yang2023Error} explored a generative error correction method using task-activating prompts, improving the error correction capabilities of ASR using zero-shot and few-shot learning. Similarly,~\cite{songspelling2023} proposed a contextual spelling correction method that optimizes LLM prompts to reduce spelling errors, particularly for out-of-vocabulary words, by adjusting ASR outputs dynamically. Both methods leverage LLMs' ability to process contextual information and improve the accuracy of the final transcription in post-processing.
The other approach is prompt-based biasing, where specific prompts are injected into LLMs to steer responses toward more accurate transcriptions based on expected terms or prior knowledge.~\cite{lei2024contextualizationasrllmusing} employed a retrieval-based method using speech similarity to provide named entities from personal databases to LLMs, reducing the error rate of named entity recognition. In a similar vein,~\cite{sun2023contextualbiasingnamedentitieslarge} used dynamic prompts combined with few-shot learning to bias the model toward recognizing rare named entities, effectively improving ASR accuracy in specialized contexts.
To reduce the cost of LLM biasing,~\cite{Flemotomos2024} proposed using vector quantization for the efficient retrieval of context, enhancing dynamic adaptation during recognition.

\section{Our method}
\label{sec:method}

\subsection{Training}

Let $\mathbf{X}$ denote the input acoustic feature sequence, which is the output of the ASR encoder in this work, and $Y = (y_1, \dots, y_T)$ be the ASR label sequence. We are also given a set of candidate phrases $\{\p_1, \dots, \p_M\}$ sampled from the transcripts within the minibatch, as well as their corresponding labels $\{l_1,\dots,l_M\}$ where $l_i=1$ if $\p_i$ is a segment of $Y$ and $l_i=0$ otherwise. Our goal is to compute a score for each phrase $\p_i$ conditioned on $\mathbf{X}$, which help distinguish the positive phrases (those with label $1$) from the negative ones (those with label $0$). We additionally introduce a special empty phrase $\p_0$, which plays the role of ``no-bias'' in other works: if all sampled phrases are distractors, we assign a label $l_0=1$ to $\p_0$, and otherwise assign $l_0=0$.

\vspace{0.5em}
\noindent\textbf{Biasing Decoder.}
We treat each phrase as a token sequence, i.e., $\p_i=\{p_{i1},\dots,p_{iL_i}\}$ where $L_i$ is the token length of $p_i$, and use an attention decoder to model the probability of a phrase $p_i$ as:
\[
P (\p_i|\mathbf{X}) = \prod_{t=1}^{L_i} P (p_{it}| \{<\!\!sos\!\!>, p_{i1}, \dots, p_{i(t-1)} \}, \mathbf{X})
\]
where $<\!\!sos\!\!>$ denotes the start of phrase and $p_{iL_i}=<\!\!eos\!\!>$ denotes the end of phrase. Note that for the empty phrase $\p_0$, the we still make one prediction on  $<\!\!eos\!\!>$. 
Essentially, the biasing decoder implements the same functionality as a normal ASR decoder, except that it models the (shorter) phrase sequences only.


\vspace{0.5em}
\noindent\textbf{Loss Functions.}  
Our training loss combines the following two components.

\begin{enumerate}
    \item \textbf{Phrase-level Log Loss:}
    We would like all positive phrases to have high probability under the biasing decoder. This is implemented by the log loss below computed over the positive phrases:
    \begin{equation*}
    \mathcal{L}_{\text{log}}
    \;=\; - \sum_{i=1}^M l_i \cdot \log P (\p_i|\mathbf{X}).
    \end{equation*}

    \item \textbf{Discriminative Loss:}  
    To ensure that true biasing phrases are strongly preferred over distractors, we introduce a discriminative loss, similar to those used for ASR~\cite{povey02,vesely2013sequence,prabhavalkar2018}.
    We first compute the phrases-level log-probabilities of both positive and negative phrases, and  them divide them by the corresponding phrase lengths to obtain averaged per-token scores:
    \begin{align} \label{eqn:per-token-score}
        s_{i} = \log P (\p_i|\mathbf{X}) / L_i .
    \end{align}
    We then normalize the resulting scores in the space of $M+1$ phrases (including the empty phrase $\p_0$), with a softmax. 
    Finally, the discriminative loss is defined as:
    \begin{equation*}
    \mathcal{L}_{\text{disc}}
    \;=\;
    - \sum_{i=0}^M l_i \cdot
    \;\log
    \frac{\exp(s_i)}
         {\sum_{j=0}^M \exp(s_j)}
    \end{equation*}
    which encourages the positive phrases to have high per-token scores relative to the negative phrases.
    
\end{enumerate}

For one training utterance, the final biasing loss \(\mathcal{L}_{\text{bias}}\) is a convex combination of the above two losses, weighted by a hyperparameter $\beta$:
\begin{equation} \label{eqn:biasing-loss}
\mathcal{L}_{\text{bias}} (X,\{\p_0,\p_1,\dots,\p_M\})
\;=\;(1 - \beta)\,\mathcal{L}_{\text{log}}
\;+\;\beta\,\mathcal{L}_{\text{disc}}.
\end{equation}
This loss is averaged over a minibatch of utterances for updating the biasing decoder. 




\begin{figure}[t]
    \centering
    \includegraphics[width=0.99\linewidth]{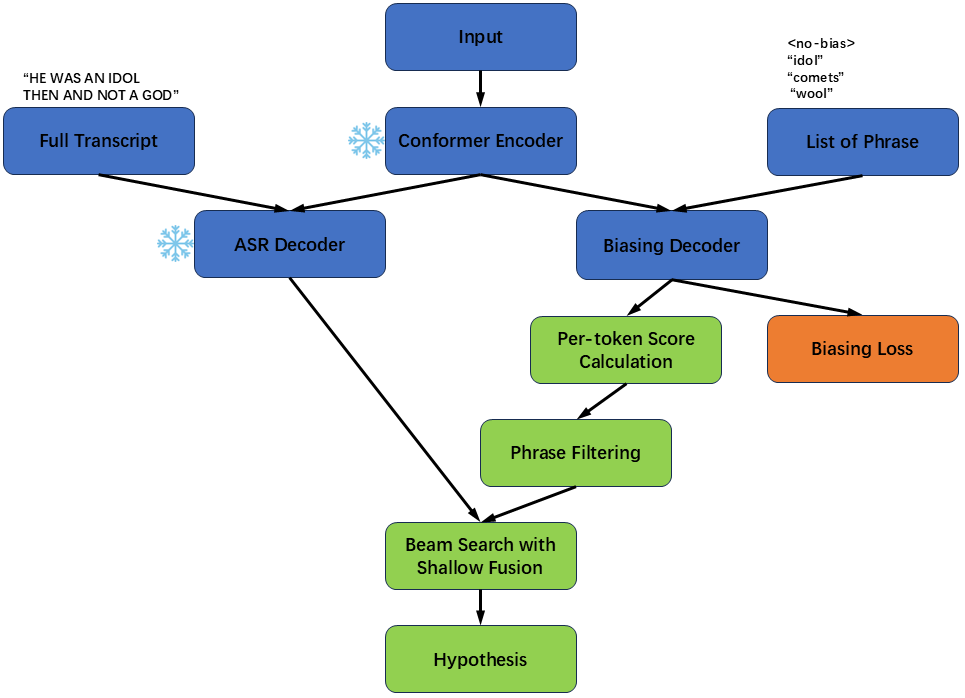}
    \caption{The overall architecture of our method.}
    \label{fig:architecture}
\end{figure}

\subsection{Inference}
\label{sec:inference}

At inference time, we perform beam search with search-based (shallow-fusion) biasing using the method of~\cite{wang2024contextual}. Note the cost of computing per-token bonus by~\cite{wang2024contextual} depends linearly on the number of phrases.
To reduce this cost, we use our neural model to perform quick filtering of the large biasing list before search starts, and compute a per-token bonus used by KMP search on the remaining phrases.    

We use the following strategy for filtering: for each phrases, we compute the per-token score as in~\eqref{eqn:per-token-score}, and compare it with the score of $\p_0$ (no-bias), we keep a phrase $\p_i$ for beam search if its per-token score is high enough so that
\begin{align}\label{eqn:keep_phrase_condition}
    tol + s_i - s_0 \ge 0 
\end{align}
for some user parameter $tol > 0$. In other words, a larger $tol$ keeps more phrases in consideration for which our model has moderate confidence. Furthermore, we use
\begin{align} \label{eqn:per-token-bonus}
  \text{bonus} = \max_i\; \{tol + s_i - s_0\}
\end{align}
as the per-token bonus for the input utterance.
Note different utterances can have different per-token bonuses.
In the initial exploration, we investigated the use of a different bonus $\{tol + s_i - s_0\}$ for each phrase (which boils down to removing $\max_i$ in~\eqref{eqn:per-token-bonus}) but obtained worse performance, so we stick to the scheme~\eqref{eqn:per-token-bonus} in this paper. 

As we will see in the experiments, the default value of $tol=0$ already works very well, and a small positive value of $tol$ can further boost the accuracy without introducing much more phrases.

An overall illustration of our method is given in Figure~\ref{fig:architecture}.

\section{Experiments}
\label{sec:expts}

\subsection{Setup}

We demonstrate our method with the Librispeech biasing setup,
which is commonly used in prior work~\cite{le2021contextual,tang2024improving,Sudo2024,Shakeel2024,Fang2025}.
The Librispeech dataset~\cite{panayotov2015librispeech} consists of 960 hours of  audio for training. The 5000 most common words in the training set accounts for 90\% of all word occurrences, and the remaining 209.2K words in the training training vocabulary are considered as \emph{rare} words. At inference time, the ASR model is supplied with a biasing phrase list containing all rare words in the reference, as well as $N=\{100, 500, 1000, 2000\}$ randomly sampled distractors (consisting of also rare words).   

\noindent \textbf{Model training.\ } Our overall model is trained in two stages. We first train the hybrid Attention-CTC ASR model with multi-task objective~\cite{watanabe2017hybrid} 
to capture general acoustic and linguistic knowledge.  Then, we freeze the ASR model and update only the biasing decoder, which is trained on bias-augmented data (with 32 phrases per utterance) with the composite loss~\eqref{eqn:biasing-loss}.
We sample biasing phrases for each training utterance on the fly. We first randomly sample 3 phrases (with 1 to 3 consecutive words) from the transcript of each utterance, to create a pool of phrases for the minibatch. Then for each utterance, we randomly sample 1 ground truth phrase from the same utterance, and 31 potential distractors from other utterances. Given that we do not filter out frequent words when sampling phrases, it is possible that phrases from other utterances end up being ground truth phrases, and we verify the phrase labels by checking each phrase against the transcript. In summary, each utterance may have multiple positive phrases as a result of the sampling process, and it is future work to avoid frequent words for phrase sampling.  

We follow the \texttt{librispeech/asr1/} recipe from Espnet~\cite{watanabe2018espnet} for ASR modeling.
The conformer encoder has 12 layers and the transformer decoder has 6 layers, both with attention dimension 512 (8 attention heads) and feed-forward dimension 2048. We perform 2x time reduction at the encoder output, by averaging every two consecutive output frames, before CTC and attention decoder. The biasing decoder has the same architecture as the ASR attention decoder. The ASR model has a total of 116M parameters and the biasing decoder has another 30M parameters. The ASR model is trained for 100 epochs, while the biasing decoder is trained for 30 epochs.

\noindent\textbf{Inference.\ } For decoding, we perform beam search with a beam size of 30. 
Before search starts, we apply the biasing decoder to filter out un-likely phrases and compute the per-token bonus as discussed in Sec~\ref{sec:inference}. Note this is a one-time cost and the model forward is done in batch mode (possibly with GPU) over the entire biasing list.

During search, The attention decoder leads the search by proposing for each hypothesis the top expansions. The log-probabilities from attention decoder are then combined with the CTC prefix scores to reduce the number of expansions to 30. Afterwards, the biasing bonuses for theses expansions are calculated and incorporated before pruning. 

\noindent\textbf{Evaluation.\ } We measure the word error rate (\textbf{WER}),  unbiased word error rate (\textbf{U-WER}), and biased word error rate (\textbf{B-
WER}) similarly to previous work.
WER is the overall word error rate measured on all words, U-WER measures the WER of words not in the biasing list, and B-WER measures the WER of words in the biasing list.
Ideally, the contextual biasing model shall have a lower B-WER without increasing its U-WER significantly. 
Furthermore, it shall also have minimal B-WER degradation as the number of distractors $N$ increases.

\begin{table}[t]
    \caption{Sensitivity with respect to $\beta$. Here $N=1000$, and $tol=0$. The average number of ground truth phrases is 2.1 for dev-clean and 1.6 for dev-other.}
    \label{tab:beta}
    \centering
    \begin{tabular}{@{}|c|@{\hspace{0.3em}}@{\hspace{0.3em}}c@{\hspace{0.3em}}|@{\hspace{0.3em}}c@{\hspace{0.3em}}||@{\hspace{0.3em}}c@{\hspace{0.3em}}|@{\hspace{0.3em}}c@{\hspace{0.3em}}|@{}}\hline
    \multirow{2}{1em}{$\beta$}  &  dev-clean & \multirow{2}{4em}{\# phrases} & dev-other & \multirow{2}{4em}{\# phrases} \\
    & WER(U-/B-WER) &  & WER(U-/B-WER) & \\
    \hline \hline
    ASR & 2.5 (1.6/9.8) &  & 6.2 (4.8/19.1) & \\ \hline\hline
    0.5 & 1.9 (1.6/4.5) & 3.0 & 5.3 (4.9/9.6) & 2.5 \\
    0.8 & 1.9 (1.6/4.6) & 2.7 & 5.3 (4.9/9.4) & 2.2 \\
    0.9 & \bf 1.9 (1.6/4.1) & 3.2 & \bf 5.2 (4.8/8.8) & 2.8 \\
    0.95 & 1.9 (1.6/4.2) & 3.4 & 5.3 (4.9/9.0) & 2.9 \\
    \hline
    \end{tabular} 
\end{table}

\subsection{Sensitivity with respect to $\beta$}
First, we perform sensitivity of the discriminative loss weight $\beta$ used in the biasing training loss~\eqref{eqn:biasing-loss}, for $N=1000$ distractors and $tol=0$.
We provide the WER/U-WER/B-WER for models trained with a range of $\beta$ values in Table~\ref{tab:beta}, as well as the number of active biasing phrases, i.e., those satisfying the condition~\eqref{eqn:keep_phrase_condition}.

Observe that for a wide range of $\beta$, the method works similarly well. The baseline ASR system achieves 9.8\% and 19.1\% B-WER for dev-clean and dev-other without biasing.
With $\beta=0.9$, we reduce the B-WERs to 4.1\% and 8.8\% respectively, without degrading the U-WERs.
Furthermore, these improvements are achieved with very small numbers of active phrases: the average number of ground truth phrases is 2.1 for dev-clean and 1.6 for dev-other, and our method only keeps on average 3.2 phrases and 2.8 phrases respectively for beam search. Therefore, our method is accurate at removing large amounts of distractors. 

Our method fails at $\beta=1.0$ (with more than 100\% WERs, not shown in the table), indicating that log loss is necessary to learn useful scores for biasing. 
From now on, we use the model trained with $\beta=0.9$.

\subsection{Sensitivity with respect to $tol$}

We have seen in Table~\ref{tab:beta} that our bonus strategy~\eqref{eqn:per-token-bonus} already works well with $tol=0$. Next we investigate if it is possible to improvement the performance with $tol>0$, at the cost of including more active phrases during search.

\begin{table}[t]
    \caption{Sensitivity with respect to $tol$. Here $N=1000$ and $\beta=0.9$. The average number of ground truth phrases is 2.1 for dev-clean and 1.6 for dev-other.}
    \label{tab:tol}
    \centering
    \begin{tabular}{@{}|c|c|c|c|c|@{}}\hline
    \multirow{2}{1em}{$tol$}  &  dev-clean & \multirow{2}{4em}{\# phrases} & dev-other & \multirow{2}{4em}{\# phrases} \\
    & WER(U-/B-WER) &  & WER(U-/B-WER) & \\
    \hline \hline
    0.0 & 1.9 (1.6/4.1) & 3.2 & 5.2 (4.8/8.8)  & 2.8 \\
    1.0 & \textbf{1.8} (1.6/3.7) & 5.5 & 5.2 (4.8/7.8)  & 5.4 \\
    2.0 & 1.9 (1.7/3.4) & 10.1 & \textbf{5.1} (4.9/7.1)  & 10.7 \\
    3.0 & 1.9 (1.7/3.1) & 17.9 & 5.2 (5.1/6.7)  & 20.2 \\
    4.0 & 2.0 (1.9/2.9) & 30.2 & 5.4 (5.3/6.4)  & 35.1 \\
    5.0 & 2.1 (2.0/2.8) & 47.9 & 5.6 (5.6/5.9)  & 56.2 \\
    \hline
    \end{tabular} 

\end{table}

In Table~\ref{tab:tol} we provide the results for several values of $tol$. Observe that indeed a positive value of $1$ or $2$ leads to improved B-WER without degrading U-WER, while the number of active phrases mildly increases from roughly 3 to about 5 and 10 respectively. Further increasing $tol$, however, tend to degrade the U-WER.

\begin{table}[t]
    \caption{Biasing with manually tuned bonus. Here $N=1000$.}
    \label{tab:manual}
    \centering
    \begin{tabular}{@{}|c|c|c|@{}}\hline
    Manual bonus  &  dev-clean & dev-other \\
    \hline \hline
    0.0 & 2.5 (1.6/9.8) & 6.2 (4.8/19.1) \\ \hline
    1.0 & 2.3 (1.6/8.0) & 5.8 (4.8/15.8) \\ \hline
    2.0 & 2.1 (1.6/6.3) & 5.5 (4.8/12.9) \\ \hline
    3.0 & 1.9 (1.6/5.1) & 5.2 (4.7/9.8) \\ \hline
    4.0 & 1.9 (1.6/4.1) & 5.1 (4.8/8.2) \\ \hline
    5.0 & \bf {1.9} (1.7/3.3) & \bf {5.0} (4.8/6.9) \\
    \hline
    6.0 & 1.9 (1.8/2.7) & 5.3 (5.2/6.0) \\
    \hline
    7.0 & 2.3 (2.3/2.5) & 5.6 (5.6/5.4) \\
    \hline
    \end{tabular} 

\end{table}

As a comparison, we exhaustively tune a constant per-token bias bonus, i.e., we replicates the method of~\cite{wang2024contextual}, and the results are given in Table~\ref{tab:manual}. We note that a carefully tuned bonus can work well for search-based biasing: with a per-token bonus of $5$, the WERs are matching the best of Table~\ref{tab:tol}. However, without the biasing decoder, we can not filter out any phrases and the number of active phrases stays at $N=1000$ during search, which is 100 times more than that of our approach.


\begin{figure}[t]
    \centering
    \includegraphics[width=0.99\linewidth]{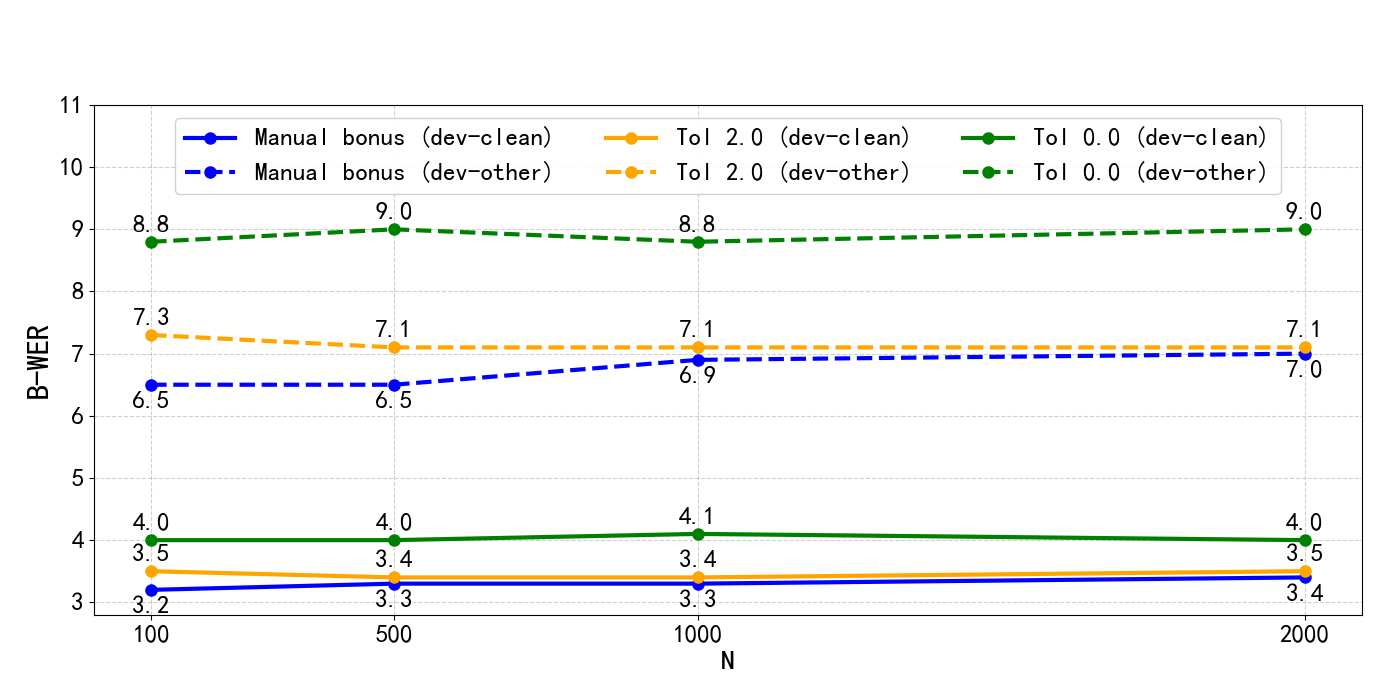}
    \caption{B-WERs of our method on dev sets, across different values of $N$.}
    \label{fig:sensitivity-n}
\end{figure}

\begin{figure}[t]
    \centering
    \includegraphics[width=0.99\linewidth]{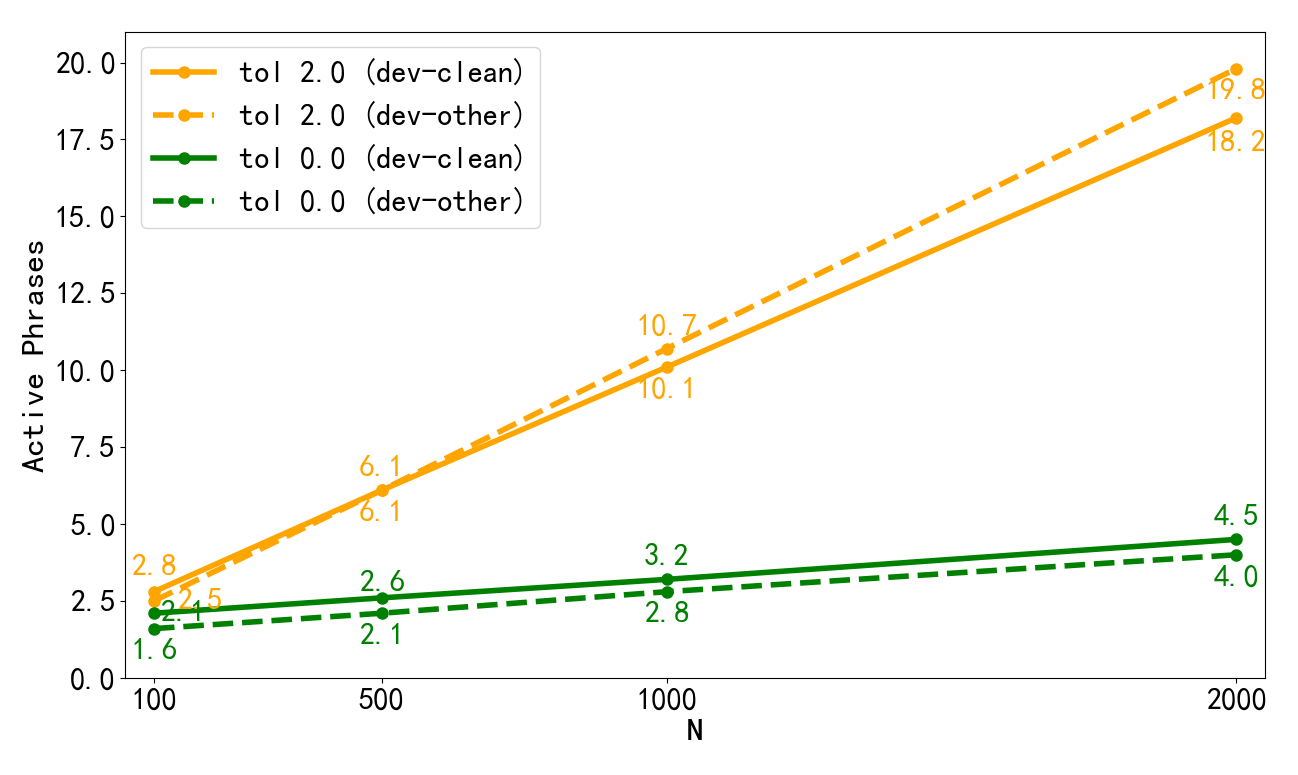}
    \caption{Number of active phrases of our method on dev sets, across different values of $N$.}
    \label{fig:num-phrases}
    
\end{figure}

\subsection{Sensitivity with respect to $N$}

Next we investigate the model's performance across a range of $N$, the number of distractors. Intuitively, as N increases, there is more chance for the model to be confused and we expect the WERs to deteriorate.

The results of our method with $tol=0$ and $tol=2$, as well as the manually tuned constant bonus 5, are shown in Figure~\ref{fig:sensitivity-n}. Since the fluctuation of WER is small, we plot only the U-WERs, which remains stable across different $N$.

In Figure~\ref{fig:num-phrases}, we plot the number of active (non-filtered) phrases in our method for varying $N$. Without biasing decoder, manual bonus cannot filter out phrases and the number of active phrases is $N$. Observe that the number active phrases grows slowly for both $tol=0$ and $tol=2$, indicating that the biasing decoder scores are consistently discriminative.

\begin{table*}[t]
    \caption{Biasing WERs of different models on Librispeech test sets. U-WER/B-WER are given in brackets. For our method, the best WERs for each $N$ are shown in boldface.}
    \label{tab:test-results}
    \centering
    \begin{tabular}{@{}|c|cccc|cccc|@{}}\hline
    \raisebox{-2ex}{Method}     &  \multicolumn{4}{|c|}{test-clean} & \multicolumn{4}{|c|}{test-other} \\
    & $N=100$ & $N=500$ & $N=1000$ & $N=2000$ & $N=100$ & $N=500$ & $N=1000$ & $N=2000$ \\
    \hline\hline

    \cite{le2021contextual}, RNN-T & \multicolumn{4}{c|}{3.7}  & \multicolumn{4}{c|}{9.6} \\ 
     & \multicolumn{4}{c|}{(2.4/14.1)} & \multicolumn{4}{c|}{(7.2/30.6)} \\ 
    DB-RNNT s3 & 2.8 & 2.9 & 3.0 & 3.0 & 8.1 & 8.3 & 8.5 & 8.8 \\
            & (2.2/7.4) & (2.3/8.1) & (2.3/8.5) & (2.3/8.9) & (7.0/17.7) & (7.1/19.1) & (7.1/20.5) & (7.3/21.8) \\
    \hline\hline

    \cite{Sudo2024}, Attention & \multicolumn{4}{c|}{5.1}  & \multicolumn{4}{c|}{8.8} \\ 
     & \multicolumn{4}{c|}{(3.9/14.1)} & \multicolumn{4}{c|}{(6.6/27.9)} \\ 
    Biasing with BPB & 2.8 & 3.2 & 3.5 & & 5.6 & 6.3 & 7.3 & \\
                & (2.3/6.0) & (2.7/7.0) & (3.0/7.7) & & (4.9/12.0) & (5.5/13.5) & (6.4/15.8) & \\
    \hline\hline

    \cite{tang2024improving}, Transducer & \multicolumn{4}{c|}{2.8}  & \multicolumn{4}{c|}{6.6} \\ 
     & \multicolumn{4}{c|}{(1.9/10.2)} & \multicolumn{4}{c|}{(4.9/22.0)} \\ 
    Biasing with CA + GA-CE & 2.2 & & 2.4 & & 5.4 & & 6.0 & \\
                & (1.8/5.1) &  & (1.9/6.4) & & (4.7/12.2) & & (5.0/15.3) & \\
    \hline \hline

    \cite{Shakeel2024}, Transducer + CTC& \multicolumn{4}{c|}{2.9}  & \multicolumn{4}{c|}{7.2} \\
    &  \multicolumn{4}{c|}{(1.6/13.0)} & \multicolumn{4}{c|}{(4.8/28.1)} \\ 
    Intermediate + Joint decoding  & 2.3 & 2.8 & 2.8 & & 6.4 & 7.1 & 7.3 & \\
              & (1.5/8.6) & (1.6/11.2) & (1.7/12.4) & & (4.7/21.2) & (4.9/26.3) & (5.0/27.9) & \\
    \hline \hline
    
    \cite{Fang2025}, Attention & \multicolumn{4}{c|}{3.4}  & \multicolumn{4}{c|}{8.8} \\
     &  \multicolumn{4}{c|}{(2.0/14.4)} & \multicolumn{4}{c|}{(6.1/32.5)} \\ 
    Biasing  & 2.3 & 2.5 & 2.7 & & 6.8 & 7.5 & 7.7 & \\
              & (1.8/6.5) & (1.9/7.4) & (2.0/8.2) & & (5.8/15.6) & (6.2/18.2) & (6.2/20.3) & \\
    \hline \hline

    \cite{Huang2024}, Transducer & \multicolumn{4}{c|}{2.2}  & \multicolumn{4}{c|}{5.2} \\
     &  \multicolumn{4}{c|}{(1.3/9.7)} & \multicolumn{4}{c|}{(3.3/21.8)} \\ 
    Shallow Fusion Biasing & 1.5 & 1.6 & 1.6 & & 4.0 & 4.1 & 4.3 & \\
    & (1.2/4.0) & (1.3/4.2) & (1.3/4.3) & & (3.3/10.5) & (3.3/11.1) & (3.5/11.2) & \\
    Neural biasing, Intermediate layers  & 1.5 & 1.7 & 1.9 & & 3.7 & 4.0 & 4.4 & \\
              & (1.1/4.7) & (1.2/5.8) & (1.3/6.6) & & (3.1/9.2) & (3.2/11.4) & (3.4/13.5) & \\
     + text perturbation  & 1.2 & 1.5 & 1.7 & & 3.3 & 3.7 & 4.2 & \\
              & (1.1/2.3) & (1.2/3.6) & (1.3/5.1) & & (3.1/5.5) & (3.2/8.2) & (3.5/10.8) & \\
    \hline \hline
    
    Ours, Attention + CTC & \multicolumn{4}{c|}{2.7}  & \multicolumn{4}{c|}{6.3} \\
    (without biasing) & \multicolumn{4}{c|}{(1.7/11.1)} & \multicolumn{4}{c|}{(4.3/23.3)}   \\ \hline
    Manual bonus (5.0) & \bf 1.9  & \bf 2.0 & \bf 2.0 &  2.2 & \bf 4.6  & \bf 4.7 & \bf 4.8 & 4.9 \\
    (no filtering)     & (1.6/4.4) & (1.7/4.6) & (1.7/4.6) & (1.8/4.9) & (4.1/9.2) & (4.2/9.4) & (4.2/9.6) & (4.4/9.9) \\ \hline
    Biasing decoder (tol 0.0) & 2.0  & 2.1  & 2.1   & \bf 2.1  & 4.9  & 4.9  & 5.0  & 5.0  \\  
                                & (1.6/5.2) & (1.7/5.2) & 1.7/5.2) & (1.7/5.3) &  (4.2/11.4) & (4.2/11.4) & (4.2/11.6) & (4.3/11.5) \\
    \hline
    Biasing decoder (tol 2.0) & 2.0  & \bf 2.0  & \bf 2.0  & \bf 2.1  & 4.7  & \bf 4.7  & \bf 4.8  & 5.0 \\  
                                & (1.6/4.8) & (1.7/4.6) & (1.7/4.6) & (1.8/4.6) & (4.1/9.6) & (4.2/9.4) & (4.3/9.3) & (4.5/9.5) \\ \hline
    Biasing decoder filtering (tol 2.0) & \bf 1.9 & \bf 2.0 & \bf 2.0 & \bf 2.1 & \bf 4.6 & \bf 4.7 & \bf 4.8 & \bf 4.8  \\
    + manual bonus (5.0) & (1.6/4.5) & (1.6/4.6) & (1.7/4.6) & (1.7/4.9) & (4.1/9.3) & (4.1/9.6) & (4.2/9.7) &  (4.2/9.7) \\
    \hline
    \end{tabular}
\end{table*}


We plot the the WERs of our method, with both $tol=0$ and $tol=2$ in Figure~\ref{fig:sensitivity-n}. We observe that the WERs stays more or less constant across N. For example, with $tol=2$, B-WER stays at roughly 3.4\% for dev-clean, and 7.1\% for dev-other. On the other hand, there is slight increase in U-WER, and this is because the model incorrectly bias towards negative phrases.
In Figure~\ref{fig:num-phrases}, we plot the number of active phrases across $N$. Even with $N=1000$ and $N=2000$, our model keeps less than 1\% of biasing list.

\subsection{Final results}

In Table~\ref{tab:test-results}, we provide the results of a few methods on the test sets.
For our method, we observe that with $tol=2$ our method performs as well as the best manually tuned bonus $5$, but we only search over a minimal amount of active phrases so that the search phase is much more efficient. We consistently achieve over 20\% relative improvement on WER, and over 50\% relative improvement on B-WER.

We also conduct an experiment where we use biasing decoder solely for filtering (with $tol=2.0$) but use manually tuned bonus $5$ for search; this illustrates the potential of using our biasing decoder for filtering while using another method for biasing.
The results are shown in the last row of Table~\ref{tab:test-results}. We achieve the best accuracy at $N=2000$, indicating that filtering not only improves the computational efficiency but also prevents over-biasing on distractors.

To put our results into context, we compare our results with those of a few recent works~\cite{le2021contextual,tang2024improving,Sudo2024,Shakeel2024,Fang2025}. 
Some of these prior works have similar baseline WERs (without biasing) to our Attention+CTC system, yet our biasing results are much stronger than theirs.
We also include the results from~\cite{Huang2024}, who achieved better WERs for the Librispeech biasing setup, with a stronger RNN-T recipe. Observe that, while their model-based biasing has an advantage for $N=100$ and $N=500$, the performance of shallow fusion remains strong, especially for $N=1000$. It is interesting future work to combine our filtering strategy with state-of-the-art model-based biasing methods.

\section{Conclusions}
\label{sec:conclusion}

We have proposed a neural model for ASR contextual biasing. Our model provides discriminative scores that can be used to effectively filter out unlikely phrases from a given biasing list, resulting in significant computational savings. When the scores are used to compute bonus for shallow-fusion biasing, they match the best manually tuned per-token bonus. 
We achieve about 50\% relative improvement in B-WER on the Librispeech biasing setup, while keep only 1\% of original phrases during search.
There are a few future directions. First, when constructing biasing phrases on the fly for training, we could remove frequent words to better simulate the scenario at testing. 
Second, we shall test our model's filtering capability with other systems that perform model-based biasing than shallow-fusion biasing, and with LLM-based ASR systems.
Third, our approach uses an ASR decoder to provide audio-conditioned biasing scores, and it would be interesting to extend it to the streaming setup.


\cleardoublepage

\bibliographystyle{IEEEtran}
\bibliography{refs}

\end{document}